\documentclass[prb,showpacs,twocolumn]{revtex4}

\usepackage[dvips]{graphicx}
\usepackage{dcolumn}
\usepackage{bm}

\begin{document}

\preprint{APS/123-QED}


\title{Pressure-induced high-$T\rm_c$ superconducting phase in FeSe: correlation between anion height and $T\rm_c$}
\author{H. Okabe$^{1}$}
 \email{okabe@phys.aoyama.ac.jp}
\author{N. Takeshita$^{2,3}$}
\author{K. Horigane$^{4}$}
\author{T. Muranaka$^{1}$}
\author{J. Akimitsu$^{1}$}%
\affiliation{
$^{1}$Department of Physics and Mathematics, Aoyama Gakuin University, Fuchinobe 5-10-1, Sagamihara, Kanagawa 229-8558, Japan\\
$^{2}$National Institute of Advanced Industrial Seience and Technology (AIST), Umezono 1-1-1, Tsukuba, Ibaraki 305-8578, Japan\\
$^{3}$JST, Transformative Research-Project on Iron Pnictides (TRIP), Tokyo 102-0075, Japan\\
$^{4}$WPI-AIMR, Tohoku University, Sendai, Miyagi 980-8577, Japan\\
}
\date{\today}
\begin{abstract}
In this study, we performed high-pressure electrical resistivity measurements of polycrystalline FeSe in the pressure range of 1-16.0 GPa at temperatures of 4-300 K. 
A precise evaluation of $T\rm_c$ from zero-resistivity temperatures revealed that $T\rm_c$ shows a slightly distorted dome-shaped curve, with maximum $T\rm_c$ (30 K) at 6 GPa, which is lower than a previously reported $T\rm_c$ value ($\sim$37 K). 
With the application of pressure, the temperature dependence of resistivity above $T\rm_c$ changes dramatically to a linear dependence; a non-Fermi-liquid-like "high-$T\rm_c$" phase appears above 3 GPa. 
We found a striking correlation between $T\rm_c$ and the Se height: the lower the Se height, the more enhanced is $T\rm_c$. 
Moreover, this relation is broadly applicable to other iron pnictides, strongly indicating that high-temperature superconductivity can appear only around the optimum anion height ($\sim$1.38$\rm\AA$). 
On the basis of these results, we suggest that the anion height should be considered as a key determining factor of $T\rm_c$ of iron-based superconductors containing various anions.
\end{abstract}

\maketitle

\section{Introduction}
Recent findings on the superconductivity in LaFeAsO$_{1-x}$F$_x$\cite{kamihara} and related materials have triggered a great deal of interest in iron compounds because of the possible connection between 
the superconductivity and magnetism,\cite{mazin} which undergoes a phase transition from antiferromagnetic to superconducting ground states (and vice versa) tuned by external pressure\cite{takahashi} or chemical doping.\cite{kamihara} 
In particular, PbO-type FeSe, which is one of the iron-based superconductors discovered a long time ago,\cite{hsu} has attracted attention as a key material for elucidating the superconducting mechanism, 
because of its extremely simple structure (composed only of the superconducting FeSe$_4$ layer) and its excellent response to external pressure.\cite{mizuguchi} 
Among all similar materials, FeSe shows the greatest enhancement of its $T\rm_c$ at high pressure:\cite{margadonna} 
$T\rm_c$ varies from 9 K (at ambient pressure) to 37 K (at 6 GPa), indicating a growth rate as high as 4.5 K/GPa; as a result, using FeSe, it is possible to demonstrate the strong correlation between the structural parameter and $T\rm_c$. 

The maximum $T\rm_c$ value of iron pnictides is apparently attained when the Fe$X_4$ ($X$: anion) tetrahedron assumes a regular shape;\cite{lee} however, this rule is not applicable to FeSe, 
because the FeSe$_4$ tetrahedron is distorted from the regular shape,\cite{margadonna} while $T\rm_c$ increases significantly with application of pressure. 
Although several studies have investigated FeSe subjected to high pressure,\cite{medvedev,garbarino} the pressure dependence of $T\rm_c$, particularly above 6 GPa, is controversial because of difficulties in measurements under high pressure conditions. 
For example, in one of these studies, superconductivity appeared to exist above 20 GPa, even though the phase transition from tetragonal to hexagonal (non-superconductive) was completed at 12.4 GPa.\cite{braithwaite} 
This discrepancy is attributable mainly to the following two reasons: the ambiguous definition of $T\rm_c$ and the large anisotropic compressibility of the layered structure.\cite{margadonna} 
FeSe does not show Meissner diamagnetism at $T\rm_c^{onset}$, which denotes the beginning of the resistivity drop; therefore, there is no guarantee that a kink in the resistivity immediately represents to a signature of superconductivity. 
Therefore, $T\rm_c$ should be decided by zero resistance temperatures. Moreover, the hydrostaticity of pressure is essential to obtain the precise pressure dependence of $T\rm_c$, 
because FeSe has an inhomogeneous compressibility,\cite{margadonna} which stems from the layered structure stacked loosely by a van der Waals interaction (see upper inset of Fig. 1). 
To overcome all these problems, we used a cubic-anvil-type high-pressure apparatus\cite{mori} that ensures quasihydrostaticity by the isotropic movement of anvil tops, 
even after the liquid pressure-transmitting medium solidifies at low temperature and high pressure; using this apparatus, we reconfirmed the $T\rm_c$-$P$ (pressure) phase diagram of FeSe. 

With this background, in this study, we measured the electrical resistivity of a high-quality FeSe polycrystal at pressures ranging from 0 GPa to 16 GPa and evaluated the variation in $T\rm_c$ and the electronic state, both of which are closely related to the anion position. 
A precise evaluation of zero-resistivity temperature shows that the pressure dependence of $T\rm_c$ has a slightly distorted dome-shaped curve with the maximum $T\rm_c$ (30 K) in the range of 0 $<$ P $<$ 11.5 (GPa) 
and that the temperature dependence of resistivities above $T\rm_c$ changes dramatically at around 2 GPa, suggesting the existence of different types of superconductivities at high pressure. 
We found a striking correlation between $T\rm_c$ and anion (selenium) height, which was obtained by a direct comparison with previous report,\cite{margadonna} $T\rm_c$ varies with the anion height. 
Moreover, this relation is broadly applicable to other ferropnictides, indicating that the high-temperature superconductivity in these materials only appears around the optimum anion height ($\sim$1.38$\rm\AA$). 
We suggest that the anion height should be considered as a key determining factor of $T\rm_c$ of iron-based superconductors containing various anions.
\section{Method}
FeSe has a simple tetragonal structure that is composed only of edge-shared FeSe$_4$ tetrahedral layers; however, it is difficult to fabricate a good-quality superconducting FeSe sample, 
because a large amount of excess iron is absolutely imperative for the occurrence of superconductivity\cite{mcqueen} and extreme caution is required to prevent the formation of magnetic impurities from easily oxidizable iron. 
Polycrystalline samples of FeSe used in this study were prepared by a solid-state reaction using Fe (99.9$\%$, Kojundo-Kagaku) and Se (99.999$\%$, Kojundo-Kagaku) powders. 
The powders were mixed in a molar ratio of 100:99 (nominal composition of FeSe$_{0.99}$) in an argon-filled glove box and sealed in an evacuated quartz tube. 
Then, the powders were sintered at 1343 K for 3 days, annealed at 693 K for 2 days, and finally quenched in liquid nitrogen. Further details of sample preparation are described in Ref. 12. 
The quality of the obtained sample was verified by powder X-ray diffraction using an X-ray diffractometer with a graphite monochromator (MultiFlex, Rigaku); 
the results confirmed that the sample quality was similar to that of previously reported high-quality samples.\cite{mcqueen} 
The electrical resistivity and magnetic susceptibility of the sample were measured using a physical property measurement system (PPMS, Quantum Design) magnetic property measurement system 
(MPMS, Quantum Design), respectively. 

As shown in Fig. 1, in our sample, zero resistivity and Meissner effect were observed simultaneously at 7 K at ambient pressure. 
In order to evaluate the precise pressure dependence of $T\rm_c$, we defined both $T\rm_c^{offset}$ (determined from the zero-resistivity temperature) and $T\rm_c^{onset}$ 
(determined from the cross point of two extrapolated lines drawn for the resistivity data around $T\rm_c$). 
Electrical resistivity measurements were performed in the cubic-anvil-type apparatus\cite{mori} with Daphne 7474 oil\cite{murata} as the liquid pressure-transmitting medium, 
which ensured precise measurements up to 16 GPa under nearly hydrostatic conditions in this study. 
Pressure was calibrated using a calibration curve that was previously obtained by observations of several fixed-pressure points (Bi, Te, Sn, ZnS) at room temperature. 
The resistivity measurements were performed by a conventional dc four-probe method, as shown in the lower inset of Fig. 1, with an excitation current of 1 mA. 
The samples used in these experiments had dimensions of 1.0$\times$0.4$\times$0.2 mm$^3$.
\begin{figure}
\includegraphics[width=1.0\linewidth,keepaspectratio]{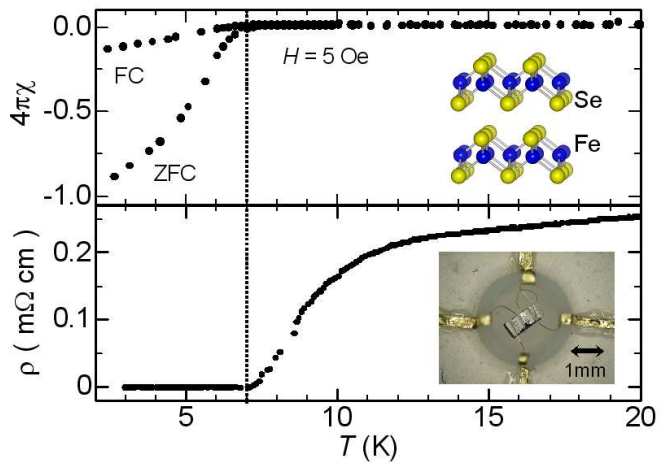}
\caption{\label{fig1}(Color online) Temperature dependence of magnetic susceptibility  (top main panel) and electrical resistivity (bottom main panel) of polycrystalline FeSe at ambient pressure. 
The top and bottom insets show the crystal structure of FeSe and the setting of the sample in the high-pressure apparatus (see text for details), respectively.}
\end{figure}
\section{RESULTS AND DISCUSSION}
The left panel of Fig. 2 shows the temperature dependence of electrical resistivities under application of external pressures ranging from 0 GPa (ambient pressure) to 16 GPa. 
With the application of pressure (ambient pressure to 16 GPa), the room-temperature resistivity decreases by a factor of more than 3; it reaches a minimum at 10 GPa and subsequently increases between 10 GPa and 16 GPa. 
In the pressure range from 0 GPa to 6 GPa, $T\rm_c$ (both $T\rm_c^{onset}$ and $T\rm_c^{offset}$) increases rapidly but not monotonically; further, the resistivity curves gradually change shape from the one at ambient pressure (see top right panel of Fig. 2). 
Meanwhile, as shown in the bottom right panel of Fig. 2, $T\rm_c^{offset}$ suddenly vanishes at 11.5 GPa; this disappearance is attributed to a rapid enhancement of resistivities between 11 GPa and 11.5 GPa. 
Although $T\rm_c^{onset}$ still remains above 11.5 GPa, it disappears completely at 16 GPa. Figure 3 shows the pressure dependence of $T\rm_c^{onset}$, $T\rm_c^{offset}$, and the width of the superconducting transition,  
$\Delta T\rm_c$ (= $T\rm_c^{onset}$ - $T\rm_c^{offset}$). Beautiful but slightly distorted dome-shaped curves are observed as cuprates\cite{palee} and heavy fermions.\cite{sidorov1} 
However, the pressure dependence of  $\Delta T\rm_c$ shows a complicated trend. At low pressures up to 2 GPa,  
$\Delta T\rm_c$ increases exponentially, indicating a salient broadening of the transition width, whereas $T\rm_c^{offset}$ increases gradually. 
Thereafter,  $\Delta T\rm_c$ decreases moderately but increases again above 9 GPa, resulting in a dome-shaped $T\rm_c$ curve. In the following paragraph, we shed light on the details of the abovementioned behaviors, 
in comparison with those reported in previous studies, to elucidate the nature of iron-based superconductors.
\begin{figure}
\includegraphics[width=1.0\linewidth]{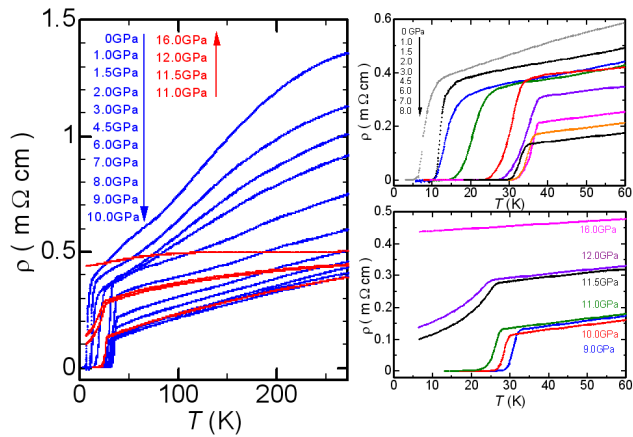}
\caption{\label{fig2}(Color online) Temperature dependence of resistivity   at ambient and several other pressures (left panel: 0$\sim$16 GPa, top right panel: 0$\sim$8 GPa, 
and bottom right panel: 9$\sim$16 GPa) for polycrystalline sample of FeSe.}
\end{figure}

The most striking feature in the low-pressure region ($<$ 2 GPa) is that $T\rm_c^{offset}$ has a relatively flat plateau; that is, an increase in $T\rm_c$ almost levels off between 1 and 1.5 GPa. 
A similar behavior was also observed during the measurements of DC magnetization\cite{miyoshi} and electrical resistivity\cite{masaki} of FeSe by using high-pressure piston-cylinder units; therefore, it is probably an important characteristic of FeSe. 
A previous $^{77}$Se-NMR measurement\cite{imai} showed that an antiferromagnetic spin fluctuation is significantly enhanced in the plateau region and that there exists a possibility of a magnetic phase transition or spin freezing. 
The superconductivity in iron-based compounds is thought to be closely related to a neighboring antiferromagnetic ordered phase; however, tetragonal FeSe exhibits superconductivity without any elemental substitution; 
therefore, its antiferromagnetic ground state remains undiscovered. Probably, in the case of FeSe, the nesting of a Fermi surface would improve temporarily up to $\sim$2 GPa by application of external pressure, 
resulting in the enhancement of antiferromagnetic instabilities such as would provide constraints on the enhancement of $T\rm_c$. 
After that, the nesting condition would worsen with the application of further pressure, and "high-temperature" superconductivity would appear. 
The appearance of pressure-induced superconductivity adjacent to a magnetic-ordered phase is a characteristic feature of exotic superconductors such as CeRh$_2$Si$_2$,\cite{movshovich} CeNi$_2$Ge$_2$,\cite{steglich} and 
CeIn$_3$,\cite{mathur} with superconductivity appearing around a quantum critical point.
\begin{figure}
\includegraphics[width=1.0\linewidth]{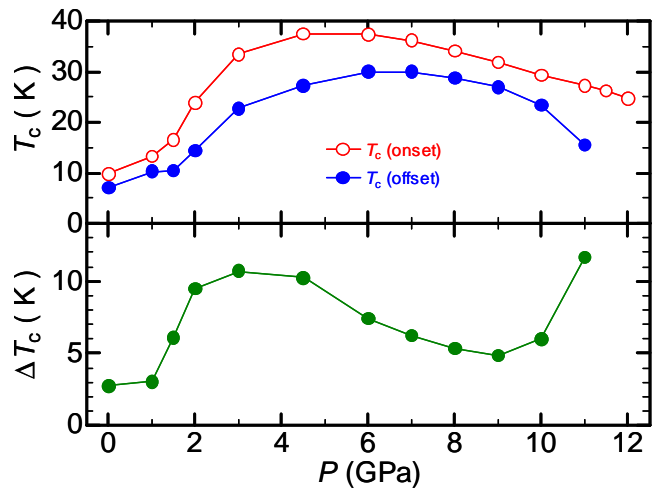}
\caption{\label{fig3}(Color online) Pressure dependence of $T\rm_c^{onset}$ (open circle), $T\rm_c^{offset}$ (closed circle, top main panel) and width of superconducting transition  
$\Delta T\rm_c$ (= $T\rm_c^{onset}$ - $T\rm_c^{offset}$) (bottom main panel). The solid lines are obtained by connecting the data points.}
\end{figure}

Figure 4 shows an enlarged view of the resistivities around $T\rm_c$ between 1 and 6 GPa. 
We can distinguish the gradual change in the shape of resistivity curves: with increasing pressure, the temperature dependence curve of resistivity changes from nearly quadratic to linear. 
In particular, the change between 2 GPa and 3 GPa is drastic, implying a phase transition between different superconducting states. 
The possibility of two kinds of superconducting phases has also been reported by Sidorov $et$ $al.$,\cite{sidorov2} indicated by the jump in $d\rho/dT$. 
A linear dependence of electrical resistivities on temperature is commonly observed in cuprate superconductors\cite{fiory,nakamura,mackenzie} and is considered to be one of the primary indicators of non-Fermi-liquid behavior; 
an incoherent scattering of fermion quasiparticles via magnetic interactions leads to resistivity of the form  $\rho (T) =  \rho_0 + AT^\alpha $   where $\rho_0$, $A$ and $\alpha$ are arbitrary constants, however, no linear term is expected according to conventional Fermi-liquid theory. 
It should be noted that in our study, the non-Fermi-liquid behavior was observed even in the plateau region (because  = 1.6$\sim$1.2 between 1 and 2 GPa), indicating the development of a spin fluctuation. 
The temperature dependence of resistivity in the high-temperature superconducting phase ($>$3 GPa) of FeSe is highly reminiscent of the linear temperature dependence observed in high-$T\rm_c$ cuprates, 
interpreted as a "strange metal" phase,\cite{anderson} where it is ascribed to antiferromagnetic spin fluctuations. 
Similar behaviors have also been reported for other ferropnictides, ex. Ba(Fe,Co)$_2$As$_2$,\cite{ahilan} implying that antiferromagnetic spin fluctuations 
and superconductivity are closely related to each other in iron-based compounds, as discussed in the context of heavy fermion and cuprate superconductors.
\begin{figure}
\includegraphics[width=1.0\linewidth]{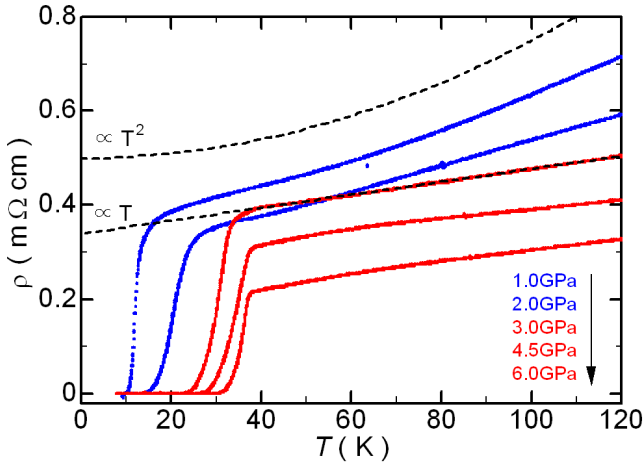}
\caption{\label{fig4}(Color online) Enlarged view of resistivity  around $T\rm_c$ between 1 and 6 GPa. The dotted lines are guides to the eye, showing the dependence of $T$ and $T^2$. For simplicity, we have not shown the data at 1.5 GPa.}
\end{figure}

For applied pressures greater than 3 GPa, $T\rm_c$ shows the dome-shaped curve, with maximum $T\rm_c$ = 30.02 K at 6 GPa, whereas between 3 GPa and 9 GPa,  $\Delta T\rm_c$ continues to decline steadily. 
As has been noted previously, the shape of the Fe$X_4$ tetrahedron is closely related to the value of $T\rm_c$; in the case of iron pnictides, $T\rm_c$ appears to attain maximum values 
when the As-Fe-As bond angles come close to 109.47$\rm^ o$,\cite{lee} which corresponds to a regular tetrahedron. However, this rule is not applicable to FeSe.\cite{margadonna} 
Therefore, we focus on the relationship of $T\rm_c$ with "Se height," which is the distance of the anion from the nearest iron layer. 
Figure 5 shows the pressure dependence of $T\rm_c^{offset}$ and Se height (inversely scaled), obtained from Ref. 6. 
Astonishingly, $T\rm_c^{offset}$ varies linearly with the Se height, even in the plateau in the low-pressure region. 
Although there is a subtle shift in the pressure dependence, which may be due to the difference in ways of applying pressures (cubic or diamond anvil), there is a clear correlation between both parameters. 
Furthermore, $T\rm_c^{offset}$ is inversely proportional to the magnitude of the Se height, as can be observed from the inset of Fig. 4, indicating that the smaller the Se height, the more enhanced is $T\rm_c$. 
However, this seems to be contradictory to the behavior observed in other pnictides: in other pnictides, it is observed that $T\rm_c$ is higher when the pnictogen is located at greater heights in the crystal structures;\cite{lee} 
this behavior is also supported by the theoretical aspect.\cite{kuroki} 
In any case, FeSe is a suitable material for demonstrating the importance of anion position as discussed below, which is inherently linked to the mechanism of superconductivity in iron-based compounds.
\begin{figure}
\includegraphics[width=1.0\linewidth]{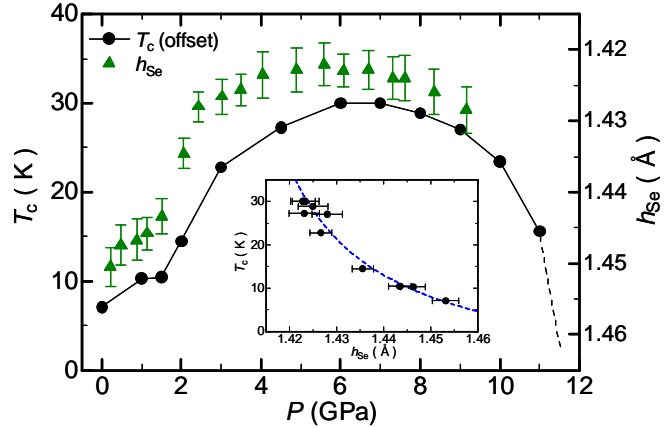}
\caption{\label{fig5}(Color online) Pressure dependence of $T\rm_c^{offset}$ and Se height $h\rm_{Se}$ (inversely scaled), as obtained from Ref. 6. The inset shows $T\rm_c^{offset}$ as a function of the Se height. The dotted line is a guide to the eye.}
\end{figure}

On application of further pressure up to 9 GPa, $T\rm_c^{offset}$ reduces monotonically to lower temperatures and disappears completely above 11.5 GPa; then, the superconducting transition becomes less sharp, 
as is indicated by the broadening of the transition width  $\Delta T\rm_c$. After the disappearance of $T\rm_c^{offset}$, the resistivity over the entire temperature range would improve greatly with increasing pressure, 
indicating the occurrence of the metal-semiconductor transition. At $\sim$9 GPa, tetragonal FeSe starts being transformed from a tetragonal structure to a hexagonal (NiAs-type) structure, 
and this structure undergoes a transition from a metallic superconducting state to the semiconducting state.\cite{margadonna} 
A recent synchrotron X-ray study on FeSe at various pressures\cite{braithwaite} has revealed that the structural transition to the hexagonal phase is completed at around 12.4 GPa, 
which is consistent with the fact that all traces of superconductivity (see bottom right panel of Fig. 2) completely vanish by 16 GPa, without any trace of an anomalous decrease in resistivity. 
Therefore, the remarkable increase in transition width  $\Delta T\rm_c$ above 9 GPa corresponds to the transition to the hexagonal phase, and this corresponds to the closure of the superconducting dome. 
The observed onset of $T\rm_c$ above 11.5 GPa indicates a subtle fraction of the superconducting phase, which may no longer manifest Meissner diamagnetism.
\begin{figure}
\includegraphics[width=1.0\linewidth]{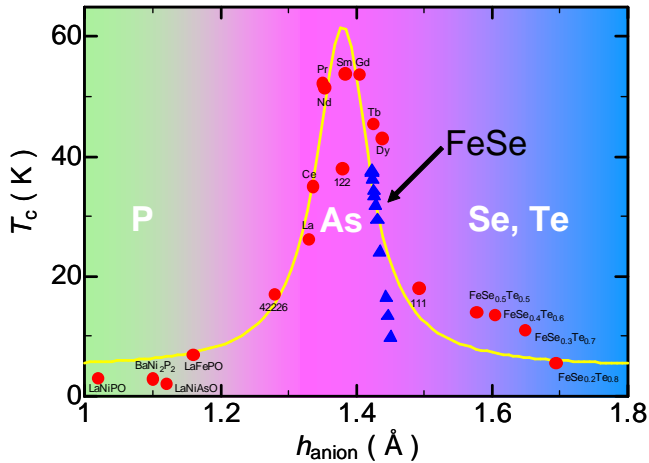}
\caption{\label{fig6}(Color online) $T\rm_c$ as a function of anion height ($h\rm_{anion}$) for various iron (and nickel)-based superconductors, as obtained from Ref. 29 (triangle: FeSe, circle: other pnictides). 
Lanthanides ($Ln$) indicate $Ln$FeAsO (1111 system). 111, 122, and 42226 represent LiFeAs, Ba$_{0.6}$K$_{0.4}$Fe$_2$As$_2$, and Sr$_4$Sc$_2$Fe$_2$P$_2$O$_6$,\cite{ogino} respectively. 
The yellow line shows the fitting result by the Lorenz function.}
\end{figure}

We now turn to consider, in a more universal sense, the nature of the iron-based superconductivity in FeSe with respect to pressure tuning of $T\rm_c$, which is the focus area in this study. 
Figure 6 shows the maximum $T\rm_c$ as a function of anion height ($h\rm_{anion}$) for various iron-based superconductors.\cite{horikin,ogino} 
In this study, we successfully derived the $T\rm_c$-$h\rm_{anion}$ diagram of iron (partially nickel)-based superconductors. 
The clear correlation between $T\rm_c$ and $h\rm_{anion}$ is a certain indicator of the importance of anion positions in these iron-based superconductors. 
As shown in Fig. 6, the anion height dependence of $T\rm_c$ is well described by a Lorenz curve (yellow line). 
As the value of anion height increases, $T\rm_c$ of the iron-based superconductors starts to increase dramatically up to $\sim$55 K at a height of 1.38$\rm\AA$, which corresponds to the optimum value of a 1111 system. 
However, above the optimum anion height (1.38$\rm\AA$), $T\rm_c$ decreases rapidly with increasing $h\rm_{anion}$, going through our measured FeSe region (1.42$\sim$1.45$\rm\AA$); 
finally, the value of $h\rm_{anion}$ becomes equal to that of non-superconducting FeTe (1.77$\rm\AA$).\cite{bao} 
It should be noted that superconductors with direct substitution in the Fe$X_4$ tetrahedral layer or a large deviation from a divalent state (Fe$^{2+}$), e.g., an alkali metal element or Co doping samples of a 122 system 
or chalcogen-substituted 11-system, are not particularly suitable for this trend. 
This is probably due to (1) the considerable disorder in the Fe layers; (2) a large gap among anion heights of different anions, for example, in FeSe$_{1-x}$Te$_x$, 
$T\rm_c$ appears to be dominated only by the Fe-Se distance ($T\rm_c$ $\sim$ 14 K at $h\rm_{anion}$ = 1.478$\rm\AA$, which is consistent with the Lorenz curve)\cite{lehman,tegel}; or 
(3) coexistence of strong magnetic fluctuation and superconductivity.\cite{lumsden1,lumsden2,parker} 
We thus conclude that the appearance of "high-temperature" superconductivity in iron compounds is confined to a specific area that is around the optimum anion height (1.38$\rm\AA$), which corresponds to the radius of arsenic at ambient pressure. 
It has been proposed,\cite{kuroki} on the basis of solutions of Eliashberg equations, that the critical temperature of iron pnictides is inherently linked to their structural parameters, particularly pnictogen heights and the $a$-axis lattice parameter. 
The result obtained in this study that the pressure evolution of $T\rm_c$ varies with the anion height, as shown in Fig. 5, is in good agreement with the theoretical prediction; 
however, the length of the $a$-axis of FeSe monotonically decreases with increasing pressure,\cite{margadonna} which suppresses the enhancement of $T\rm_c$. 
An interesting aspect of FeSe, as observed from Fig. 6, is that $T\rm_c$ does not exhibit this trend above 1.43$\rm\AA$ (corresponding to the pressure range of 0$\sim$2 GPa), 
which clearly indicates that the system attains a different electronic state below the characteristic pressure ($\sim$2 GPa). 
This could occur concomitantly with the reconstruction of a Fermi surface; the shapes of the resistivity curves above $T\rm_c$ change clearly at around 2 GPa, as pointed out above (see Fig. 4), 
which implies a significant transformation to the non-Fermi-liquid state. It has been previously suggested that there is a difference in the superconducting gap symmetries of arsenic and phosphide:\cite{fletcher} 
a full-gap strong coupling $s$-wave for high-$T\rm_c$ arsenide compounds and nodal low-$T\rm_c$ for phosphide compounds, which is widely perceived in many studies. 
A theoretical approach\cite{kuroki} has suggested that the pairing symmetry of iron pnictides is determined by the pnictogen heights between a high-$T\rm_c$ nodeless gap for high $h\rm_{anion}$ or a low-$T\rm_c$ nodal gap for low $h\rm_{anion}$, corresponding to the left-hand side of the Lorenz curve shown in Fig. 6. 
Although FeSe is located on the right-hand side, i.e., in a region of extremely high $h\rm_{anion}$, it is highly probable that FeSe shows two or more different types of superconductivities under application of external pressures, 
as is the case with pnictides. The extremely soft crystal structure of FeSe enables the control of $h\rm_{anion}$ in a wide range and the superconducting mechanism can be switched by the application of modest pressure. 
It may be interesting to explore the gap symmetry of FeSe at high pressure ($\sim$6 GPa) by NMR or muon spin rotation and whether there is any difference between the gap symmetry of FeSe and those of other iron-based superconductors.
\section{SUMMARY}
In this study, the precise pressure dependence of the electric resistivity of FeSe was measured in the pressure range of 0-16.0 GPa at temperatures of 4-300 K by using a cubic-anvil-type high-pressure apparatus. 
$T\rm_c$ estimated from zero-resistivity temperature shows a slightly distorted dome-shaped curve, with the maximum $T\rm_c$ = 30 K in the range of 0 $< P <$ 11.5 (GPa), which is lower than those in previous studies. 
The temperature dependence of resistivity above $T\rm_c$ changes dramatically at around 2$\sim$3 GPa; the shapes of the resistivity curves change to linear shapes, which is one of the primary indicators of non-Fermi-liquid behavior; 
this behavior strongly suggests a phase transition between different superconducting states. A striking correlation is found between $T\rm_c$ and anion (selenium) height: the lower the Se height, the more improved is $T\rm_c$. 
Moreover, this relation is broadly applicable to other iron pnictides, indicating that the high-temperature superconductivity in these materials appears only around the optimum anion height ($\sim$1.38$\rm\AA$). 
On the basis of these results, we suggest that anion height should be considered as a key determining factor of $T\rm_c$ in iron-based superconductors containing various anions.

\vspace{5mm}
We would like to acknowledge Dr. K. Kuroki (The Univ. of Electro-Communications) for helpful discussions. This work was supported by "High-Tech Research Center" 
Project for Private University from the Ministry of Education, Culture, Sports Science and Technology (MEXT), 
Grant-in-Aid for Yong Scientists (B) (No. 20740202) and a Grant-in-Aid for Specially Promoted Research from MEXT, and the Asahi Glass Foundation.

\end{document}